\documentclass[pre,twocolumn]{revtex4}
\usepackage[english]{babel}
\usepackage{amsmath}
\usepackage{amssymb}
\usepackage{epsfig}

\begin{document}
\title{``Capillary'' structures in transversely trapped nonlinear optical beams}
\author{Victor P. Ruban}
\email{ruban@itp.ac.ru}
\affiliation{Landau Institute for Theoretical Physics RAS,
Chernogolovka, Moscow region, 142432 Russia}

\date{\today}

\begin{abstract}
A mathematical analogy between paraxial optics with two circular polarizations 
of light in a defocusing Kerr medium with positive dispersion, binary Bose-Einstein
condensates of cold atoms in the phase separation regime, and hydrodynamics of two
immiscible compressible liquids can help in theoretical search for unknown 
three-dimensional coherent optical structures. In this work, transversely trapped
(by a smooth profile of the refractive index) light beams are considered and 
new numerical examples are presented, including a ``floating drop'', a precessing
longitudinal optical vortex with an inhomogeneous profile of filling with the
second component, and the combination of a drop and a vortex filament. Filled
vortices that are perpendicular to the beam axis and propagate at large distances
have also been simulated.

\vspace{1mm}

\noindent V. P. Ruban, JETP Lett. {\bf 117}(4), 292 (2023); 
DOI: 10.1134/S0021364022603311
\end{abstract}

\maketitle

\subsection*{Introduction}

As known, equations of paraxial optics describing
the propagation of two interacting circular polarizations 
of light in a dielectric medium with Kerr nonlinearity [1-4] 
in the completely defocusing case are mathematically 
equivalent to the coupled Gross-Pitaevskii equations 
for a binary Bose-Einstein condensate in
the spatial phase separation regime [5-11]. In turn,
the behavior of Bose-Einstein condensates at relatively 
large scales in many respects corresponds to
classical hydrodynamics. A number of phenomena
similar to the behavior of classical immiscible liquids
were theoretically revealed in the physics of ultracold
gas mixtures. In particular, the effective surface tension
[8, 12] is responsible for ``capillary'' phenomena
such as bubble dynamics [13], analogs of classical
hydrodynamic instabilities (Kelvin-Helmholtz [14-16], 
Rayleigh-Taylor [17-19], and Plateau-Rayleigh [20]), 
parametric instability of capillary waves at the
interface [21, 22], complex textures in rotating binary
condensates [23-25], three-dimensional topological
structures [26-29], capillary buoyancy of droplets in
trapped immiscible Bose-Einstein condensates [30],
and vortices with a filled core [7, 31-38]. Many of
these phenomena should occur in nonlinear optics,
but some of them have not yet been observed in optical
laboratories. However, at least domains of polarization
of light have been known for a long time (see [39, 40]
for counterpropagating waves and [41-45] for 
codirectional waves).

The concept of surface tension allows the qualitative 
analysis of many such structures, which is particularly
valuable for three-dimensional configurations.
This concept corresponds to the regime of extremely
strong nonlinearity and is sometimes an indispensable
tool for study besides the hydrodynamic approximation. 
In this work, using recent achievements in the
field of cold gases and the common intuitive understanding 
of the properties of capillarity based on daily
practice, some important aspects of filled optical vortices, 
which are known two-component objects, were
examined [43-47]. Furthermore, three likely novel
capillary structures in transversely trapped optical
beams --- simple floating drop, precessing floating drop
with longitudinal vortex filaments attached to it, and
filled vortices perpendicular to the beam axis --- are
``constructed'' numerically.

\subsection*{Model}

A fundamentally insignificant difference of optics
from hydrodynamics is that the evolutionary variable
in optics is usually the distance $\zeta$ along the beam axis
rather than the time $t$, and the ``delayed'' time $\tau=t-\zeta/v_{\rm gr}$
serves as the third ``spatial'' coordinate. For this
reason, some effort is sometimes required to realize
the four-dimensional picture of a phenomenon. In particular, 
this remark concerns the motion of quantized vortices, which 
can have different orientations in the $(x,y,\tau)$ space.

We consider a dielectric medium with an isotropic
dispersion relation of linear optical waves
$k(\omega)=\sqrt{\varepsilon(\omega)}\omega/c$
and with defocusing Kerr nonlinearity. Let the dispersion
of the group velocity be anomalous (i.e., $k''(\omega)<0$ in a certain range). 
We fix the carrier frequency $\omega$, the corresponding wavenumber $k_0$,
and the second derivative $k_0''$. We also assume that the
part of the permittivity linear in the electric field in the
spatial region of interest is slightly inhomogeneous
and has an approximately parabolic two-dimensional
profile at the frequency $\omega$:
\begin{equation}
\varepsilon_{\rm lin}=\varepsilon[1+\nu^2(2-(x^2+\kappa^2 y^2)/R^2)],
\end{equation}
where $\nu\ll 1$ is a small dimensionless parameter,
$R\gg 1/k_0$ is the relatively large length parameter
(characteristic width of the light beam), and $\kappa$ is the
coefficient specifying the transverse geometric anisotropy of
such a smooth waveguide (for certainty, $\kappa\geq 1$). 
A similar medium was considered, e.g., in [48] (for one
polarization of light).

Below, the optical beam with both circular polarizations
is considered. In terms of the appropriate dimensionless variables,
the system of equations for slow complex envelopes 
$A_{1,2}(x,y,\tau,\zeta)$ corresponding to the right and left
circular polarizations of light has the form
\begin{equation}
i\frac{\partial A_{1,2}}{\partial \zeta}=\Big[-\frac{1}{2}\Delta 
+V(x,y)+|A_{1,2}|^2+ g_{12}|A_{2,1}|^2\Big]A_{1,2},
\label{A_12_eqs}
\end{equation}
where $\Delta=\partial_x^2+\partial_y^2+\partial_\tau^2$ is the 
three-dimensional Laplace operator in the $(x,y,\tau)$ ``coordinate'' space.
These equations are easily derived by analogy with the
section ``Self-focusing'' in L.D. Landau and E.M. Lifshitz, 
{\it Electrodynamics of Continuous Media} (Pergamon, Oxford, 1984). 
To this end, the expression for the electric field in terms of
circular polarizations
\begin{equation}
{\bf E}=\big[({\bf e}_x+i{\bf e}_y) A_1 
           + ({\bf e}_x-i{\bf e}_y) A_2 \big]/\sqrt{2}
\end{equation}
should be substituted into the formula
\begin{equation}
{\bf D}^{(3)}=\alpha(\omega)|{\bf E}|^2{\bf E}+\beta(\omega){\bf E}^2{\bf E}^*,
\end{equation}
the remaining algebra should be performed (taking into account 
the second time derivative), and the following rescaling should be done:
$\zeta=(\nu/R)\cdot\zeta_{\rm old}$, 
$\{x,y\}=\sqrt{\nu k_0/R}\cdot \{x,y\}_{\rm old}$,
$\tau=\sqrt{\nu/(R|k_0''|)}\cdot\tau_{\rm old}$, and 
$A_{1,2}=\sqrt{k_0 R|\alpha|/(2\nu \varepsilon)}\cdot A_{1,2}^{(\rm old)}$. 

It is noteworthy that the corresponding equations
obtained in the basis of linearly or elliptically polarized
waves will include additional terms with four-wave
mixing, which is less convenient for analysis. Nonlinearity 
is defocusing if $\alpha$  and $\beta$ are negative. The cross-phase 
modulation parameter $g_{12}=1+2\beta/\alpha$ depends
on the material; in a typical case, it is 2. The defocusing 
character of nonlinearity and the condition of
strong cross repulsion $g_{12} > 1$ imply the phase separation 
regime. In other words, domains with opposite
circular polarizations are spontaneously formed at
intensities of light $|A_{1,2}|^2\sim I$ in the $(x,y,\tau)$ space at
scales $l\gg 1/\sqrt{I}$ [41-45]. The initial stage of this process 
is often a specific modulation instability of states
with linear or elliptic polarization. The geometric
shape of these domains changes generally with
increasing $\zeta$, which corresponds to time evolution in
hydrodynamics. The thickness $w\sim 1/\sqrt{I}$ of domain
walls in this case is about the inverse typical amplitude
and the surface tension is $\sigma\sim I^{3/2}$.

The effective external potential $V(x,y)$ in Eqs. (2) is proportional
to the deviation $\tilde \varepsilon(x,y)$ of the permittivity
from a constant value taken with the minus sign. In our case,
\begin{equation}
V(x,y)=(x^2+\kappa^2 y^2)/2 -\mu, \qquad \mu=\nu k_0 R.
\label{V_xy}
\end{equation}
Inhomogeneity $\tilde \varepsilon$ is necessary for the transverse
confinement of the optical beam undergoing diffraction
and nonlinear defocusing. The refractive index can generally 
depend on $\zeta$, but this dependence should be sufficiently
slow for the physical applicability of the approximation 
under consideration. For the validity of analogy with 
optics in the case of Bose-Einstein condensates, 
the potential of the trap should depend on
no more than two Cartesian space coordinates and
possibly on the time. Below, we focus on the purely
two-dimensional potential ``valley'' given by Eq. (5).
In ``equilibrium,'' when the functions $A_{1}$ and $A_{2}$ are
independent of $\zeta$, the length of the light beam in the
variable $\tau$ is thus infinite. This property implies that
possible two-dimensional solutions (independent of $\tau$)
should be unstable because a stable longitudinal
domain wall can hardly be realized in this situation;
domains with transverse walls alternating along the $\tau$ 
axis seem more realistic. At the same time, the increment 
of longitudinal instability will be very small for
long-wavelength (i.e., low-frequency) perturbations, and their
amplitudes will increase strongly only at fairly large $\zeta$
values when many events occur in two-dimensional dynamics.

The parameter $\mu$  (in the theory of Bose-Einstein
condensates, it is the dimensionless chemical potential) 
will characterize the level of nonlinearity of the
background state of the optical field ($I_1\sim \mu$). It is
assumed to be large enough for the transverse dimension 
of the beam $R_\perp \sim \sqrt{\mu}$ to be much larger than the
characteristic thickness of domain walls $w$ and the core
$\xi$ of quantized vortices: $w\sim\xi\sim 1/\sqrt{\mu}$. At the same
time, overly high powers can be undesirable in practice. 
For this reason, the moderate value $\mu=6.0$ is used in numerical examples.

The parameter $\nu$  should be small to ensure the
applicability of the quasimonochromatic approximation 
because the characteristic width of vortices and
domain walls in the initial dimensional variables is
estimated as $w_\perp\sim (\nu k_0)^{-1}\gg 1/k_0$ in the transverse
coordinates and as $w_\tau\sim \sqrt{|k_0''|/k_0}/\nu$ in the time coordinate. 
Therefore, (anomalous) dispersion should not be too 
weak to ensure the condition $w_\tau\gg 1/\omega$ .

Realistic parameters seem to be $\nu\sim 0.03$ and
$k_0R\sim 200$, which constitutes about 30 wavelengths
(then, about five wavelengths in the thickness of vortices 
and domain walls). The characteristic distance
along the beam $R/\nu$ is about a thousand wavelengths,
i.e., about a millimeter. The $\zeta$ interval of several hundred 
units corresponds to tens of centimeters. The system 
under consideration is in principle similar to
graded-index multimode optical fibers, although light
beams in the latter case are narrower, only a few wavelengths 
(see, e.g., [49-53] and references therein). A
more important difference is the defocusing character
of nonlinearity in our case, whereas conventional optical 
fibers are made of materials with focusing nonlinearity. 
For this reason, a fast experimental test of predictions
of the developed theory is hardly possible
because special efforts are required to fabricate a sample 
with the necessary properties. It is important that
the defocusing Kerr dielectric with anomalous dispersion
and a smooth profile of the refractive index is not
fundamentally forbidden (see, e.g., [48]). Since
numerical solutions for such systems are very interesting, 
these theoretical results need serious attention.

\begin{figure}
\begin{center}
\epsfig{file=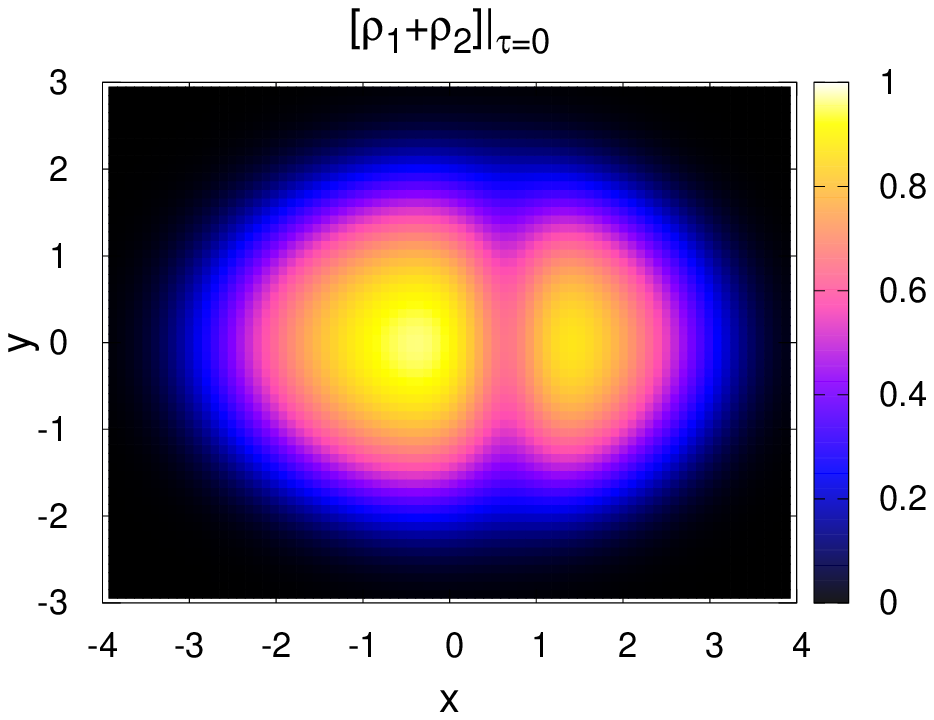, width=54mm}\\
\epsfig{file=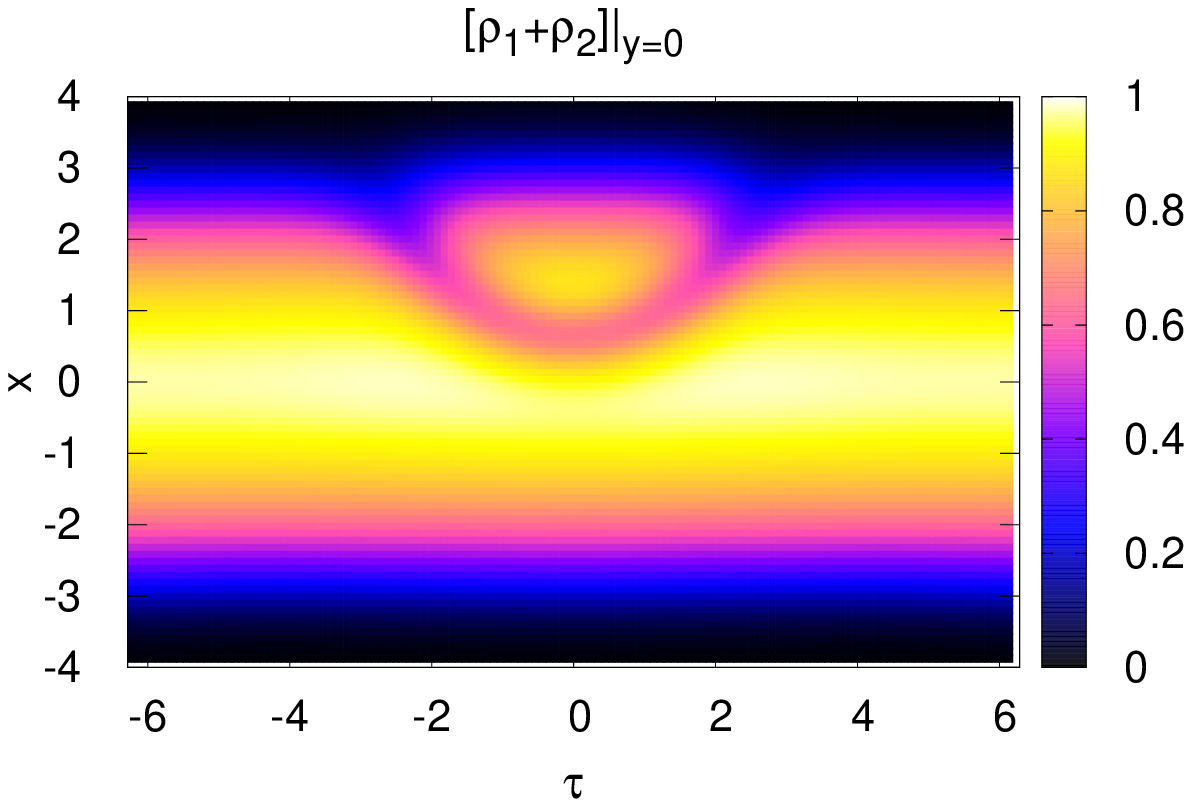, width=75mm}
\end{center}
\caption{Numerical example of the light beam with the floating drop
at the variable $\zeta=300$. Color represents the total normalized
density $\rho_1+\rho_2=(I_1+I_2)/\mu$ (a) in the cross section of the beam
by the $\tau=0$ plane and (b) in the longitudinal section of the
beam by the $y=0$ plane.
}
\label{floating_drop} 
\end{figure}

\begin{figure}
\begin{center}
\epsfig{file=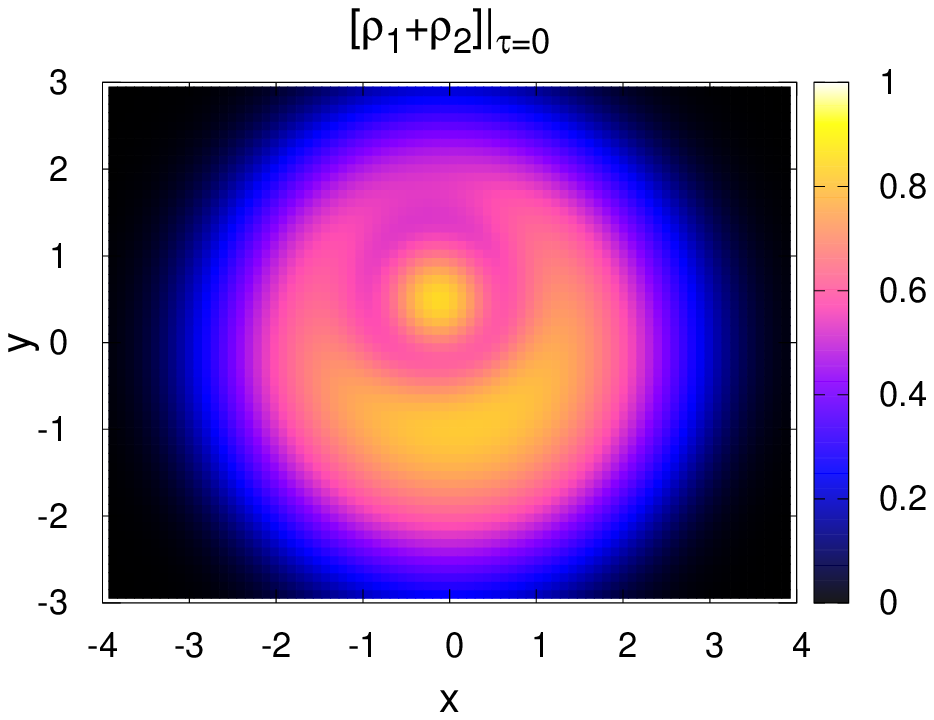, width=54mm}\\
\epsfig{file=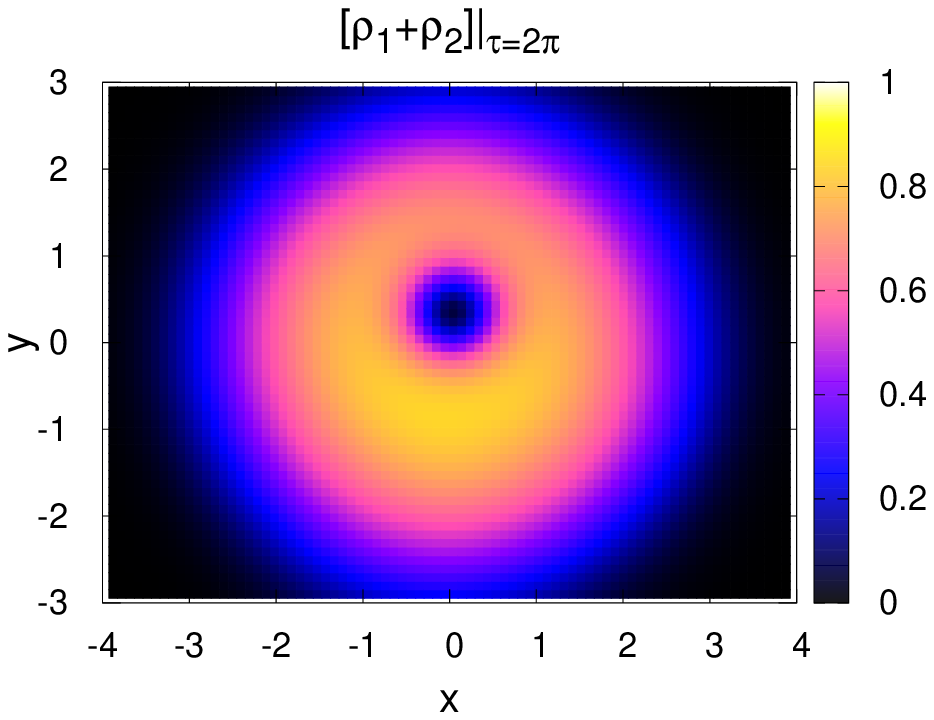, width=54mm}\\
\epsfig{file=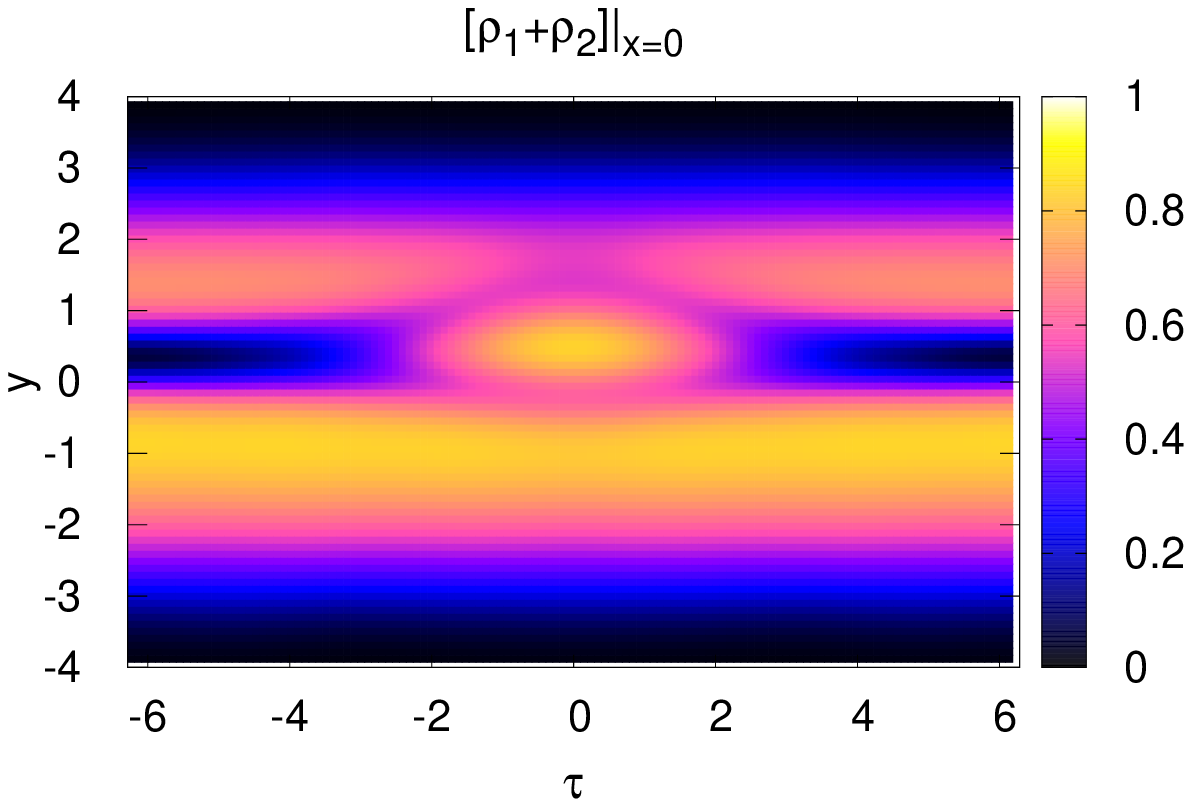, width=75mm}
\end{center}
\caption{ Light beam with the precessing vortex-soliton
complex at $\kappa=1$ and $\zeta=228$: (a) the cross
section of the beam by the $\tau=0$ plane, (b) the cross section
of the beam by the $\tau=2\pi$ plane, and (c) the longitudinal 
section of the beam by the $x=0$ plane.
}
\label{SV} 
\end{figure}

\subsection*{Numerical method}

The system of Eqs. (2) was solved numerically in
the cubic region $(4\pi)^3\approx (12.6)^3$ with periodic boundary 
conditions in variables $(x,y,\tau)$ by the second-order
Fourier method with a split step in the variable $\zeta$. The
accuracy of calculations was controlled by the conservation 
of the Hamiltonian of the system to the sixth decimal place. 
Nearly steady states at $\zeta=0$ were prepared 
by the imaginary-time propagation method. In
other words, $i$ was replaced by $-1$ on the left-hand
sides of Eqs. (2) and this dissipative dynamics was calculated 
in a certain finite pseudotime interval. In this
case, a certain positive correction $\delta \mu_2$ was added to the
chemical potential of the second component. This
correction and bare profiles of both wave amplitudes
were selected using the trial-and-error method to
obtain the desired ``mass'' of the second component.
This dissipative procedure strongly suppressed hard
excitations, whereas soft degrees of freedom of interest
remained unrelaxed. After that, the calculation of conservative
system (2) began.

Several substantial numerical examples are given below.

\subsection*{Floating drop}

We begin with the simplest case where almost the
entire beam (with the transverse anisotropy parameter
$\kappa^2=2$ ) consists of the first component where a drop
of the second component floats (see Fig. 1, where the
domain wall is seen as a dip of the total intensity).
Quantized vortices are absent. Imaginary-time relaxation 
at the stage of preparation of the initial state leads
to a nearly static initial configuration. The shape of the
drop slightly oscillates with the conditional ``time'' $\zeta$.
Such a behavior is intuitively clear because we deal
with a natural mechanical system where the minimum
of the potential energy is reached at given total masses
of both components and small oscillations near this
minimum occur. However, an overly massive drop is
unstable and, as a result, the drop fills the entire section 
of the beam in a certain part of it (such a simple
quasi-one-dimensional structure is not shown here).

\begin{figure}
\begin{center}
\epsfig{file=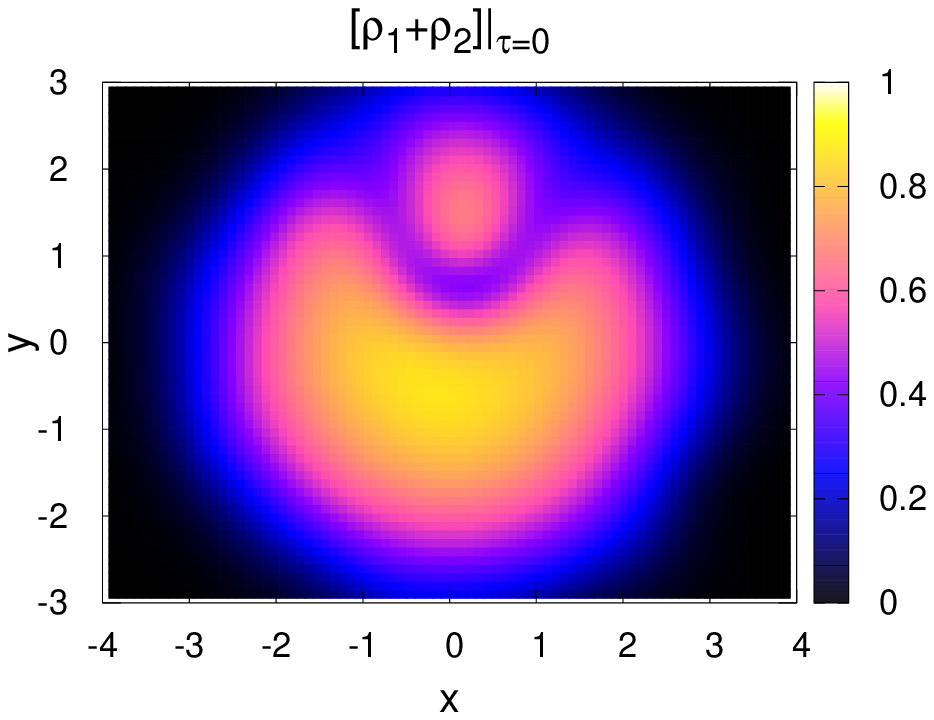, width=54mm}\\
\epsfig{file=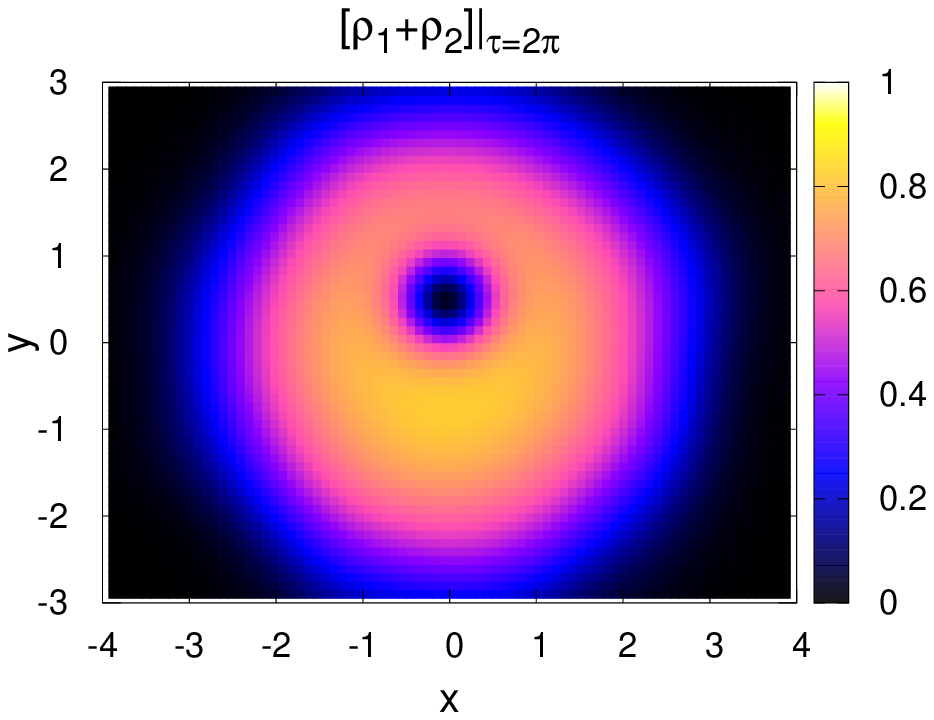, width=54mm}\\
\epsfig{file=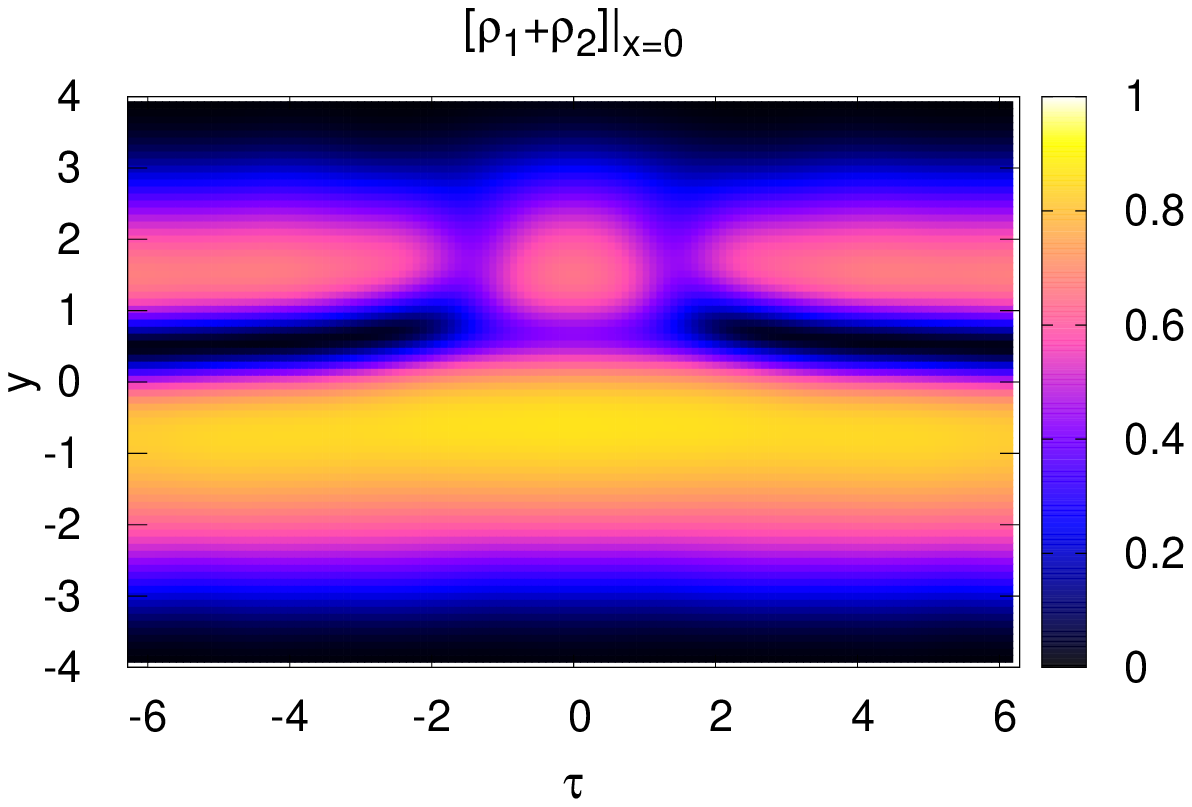, width=75mm}
\end{center}
\caption{Precessing floating drop with
attached vortex filaments at $\kappa=1$ and $\zeta=253$: (a) the
cross section of the beam by the $\tau=0$ plane, (b) the cross
section of the beam by the $\tau=2\pi$ plane, and (c) the longitudinal 
section of the beam by the $x=0$ plane.
}
\label{DV} 
\end{figure}

\subsection*{Longitudinal vortex with inhomogeneous filling}

Very interesting numerical solutions are obtained if
the confining potential is axisymmetric ($\kappa =1$) and a
longitudinally oriented (single) optical vortex with a
filled core is located at a small distance from the beam
axis at $\zeta=0$ . In the presence of small initial inhomogeneity,
the so-called sausage instability is developed;
it is caused by surface tension and results in the formation
of a spindle-shaped bubble with attached input
and output vortex filaments (this phenomenon was
discussed in [36]). In the case of a moderate amount
of the second component, the formation and decay of
the bubble can be repeated several times. In this case,
the vortex precesses about the beam axis. The vortex
with a sufficiently appropriate inhomogeneous filling
can immediately be initiated; in this case, the bubble
approximately keeps its shape, rotating about the
beam axis (see video [54], where the dynamics of the
conditional surface determining the core of the vortex
is shown; color corresponds to the $x$ coordinate).
Thus, the sausage instability is saturated and a precessing
three-dimensional vortex-soliton complex is formed, 
exemplified in Fig. 2. It is seen that the length
of the second component along the beam is finite: it is
entirely ``extruded'' by the surface tension from the
remaining part of the vortex to the bubble. Since this
complex in the numerical experiment expands to hundreds
of units in the variable $\zeta$ and is held under
``unsteady'' perturbations, it is assumingly stable.
However, this problem requires further study. This
example is a nontrivial generalization of the structure
previously known for a uniform medium (without the
confining external potential) [43–47].

It should be mentioned that, if a certain longitudinal 
``velocity'' $u$ is assigned to the second component
at $\zeta=0$ (multiplying $A_2$ by $\exp(i u\tau)$, which really
means a small reduction of the frequency of the second 
component), the soliton becomes moving and the entire structure
becomes slightly twisted (see video [55]).

\begin{figure}
\begin{center}
\epsfig{file=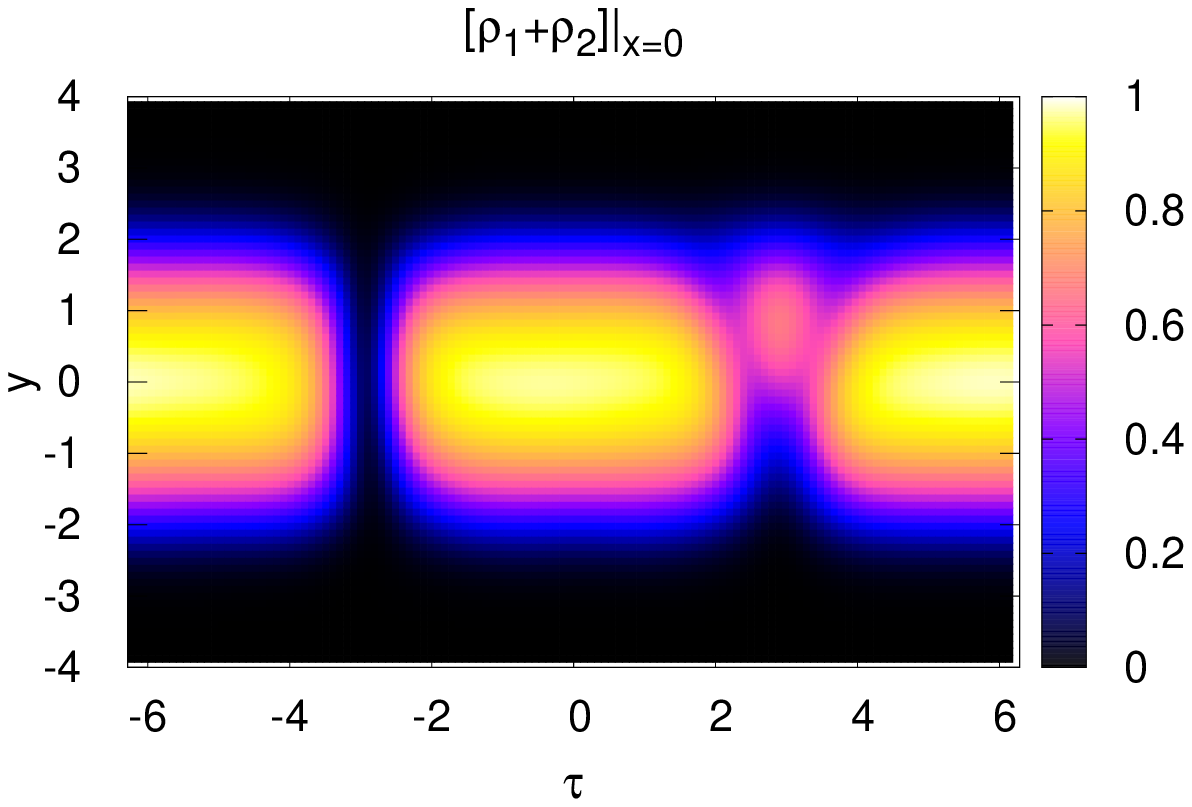, width=75mm}\\
\epsfig{file=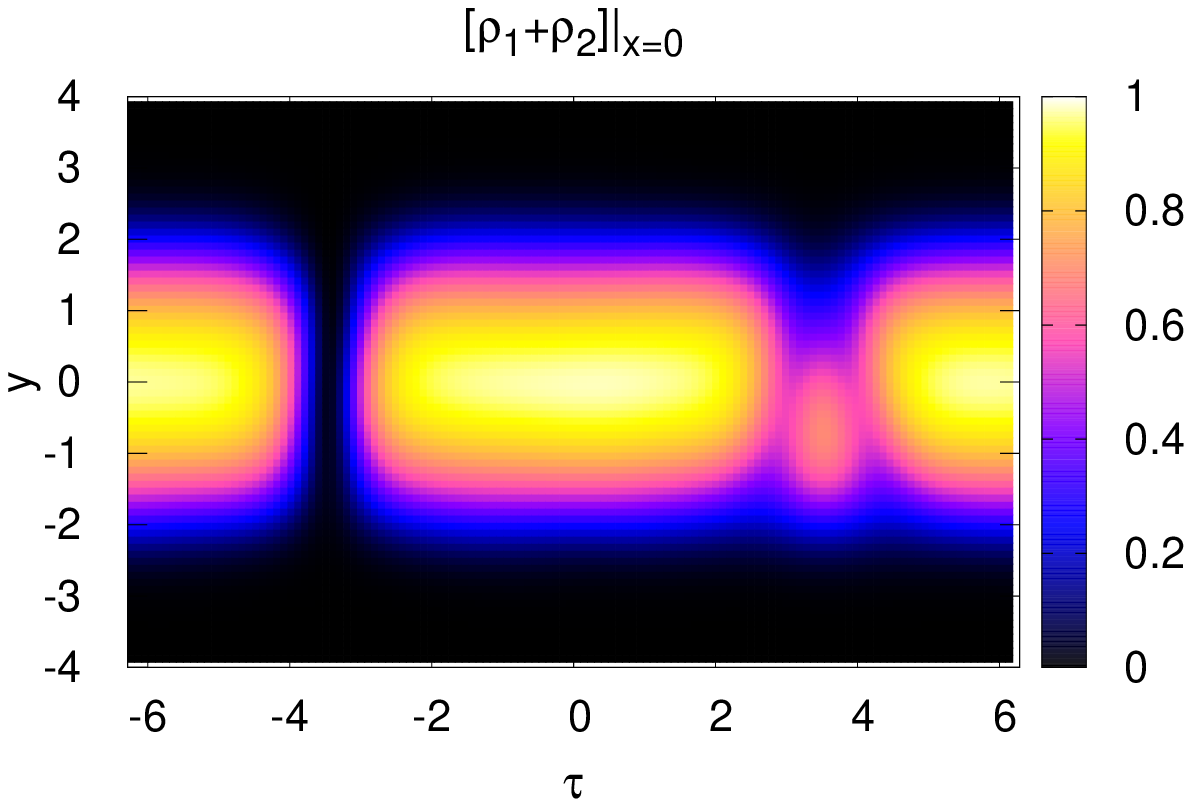, width=75mm}
\end{center}
\caption{System of two transverse vortices at $\kappa^2=2$. 
The positive and negative vortices are oriented
along the $y$ axis and their cores differ in the amount of the
trapped second component. The longitudinal section of
the beam by the $x=0$ plane at $\zeta =$ (a) 291 and (b) 300.
}
\label{perp} 
\end{figure}

\subsection*{Floating drop with attached vortex filaments}

The scenario of evolution at the quite large filling
of the initial vortex is different. In this case, a massive
lemon-shaped bubble is formed on the vortex and,
then, its precession becomes unstable. The bubble
floats to the surface of the beam and is transformed to
a floating drop with two empty vortex filaments
attached to it. After transient processes, the system
again undergoes almost stationary precession about
the beam axis (see video [56]). The corresponding
example is shown in Fig. 3. Such a complex solution
obviously cannot be obtained by any existing analytical 
methods. However, being constructed numerically,
this solution seems natural and convincing.

\subsection*{Transverse vortices with filling}

Finally, we examine the possibility that vortices in
the general case can be oriented arbitrarily in the
$(x,y,\tau)$ space, in particular, across the beam. The
calculations showed that such transverse configurations
are long-lived. Figure 4 presents the situation with two
opposite transverse vortices; the second vortex is necessary 
to satisfy the periodic boundary condition in the variable 
$\tau$. One of the vortices is filled and the second 
is almost empty. The distance between the vortices
in this numerical experiment was periodically varied
but without their close approach (see video [57]).
Moreover, since the center of masses of the second
component was first above the $y=0$ plane, slow
vibrations of the second component along the $y$ axis
occur in the process of propagation. These vertical
vibrations are seen in Fig. 4.

Dynamics in the presence of a larger number of such vortices
has not yet been simulated because this simulation would require
a sufficiently long computational region in the variable $\tau$. 
This problem remains for future studies.

It is interesting to note that a sequence of such
transverse vortices can carry information because the
filling of each vortex holds for a long time and can be
used for encoding. It is still unclear whether this property 
can be used in practice.

\subsection*{Conclusions}

To summarize, the concept of surface tension has been applied 
to nonlinear and unsteady optical beams with both circular polarizations. 
A number of new significantly three-dimensional combined coherent
structures caused by phase separation have been found.

This field of research seems quite promising
because almost any reasonable initial configuration
leads to interesting subsequent dynamics. In particular, 
the simulation of the interaction of two floating drops coupled
by a vortex filament, as well as the consideration of 
the dynamics of polarization domains in the presence of transverse
vortices against a significantly nonparabolic background intensity
profile [e.g., in the two-well potential $V(x,y)$], remains for the future.

It is noteworthy that the interaction between two polarizations 
leads to interesting structures in systems with more complex nonlinearity
as well (see, e.g., [58, 59]).

A certain time is required to verify whether the predicted 
solutions will be implemented experimentally.
As mentioned above, this is hardly possible with existing 
samples, and appropriate experimental materials
should be prepared from the ``beginning''.

\subsection*{Acknowledgments}

I am grateful to E.A. Kuznetsov for a valuable remark
initiating this study.

\subsection*{Funding}

This work was supported by the Ministry of Science and
Higher Education of the Russian Federation (state assignment 
no. 0029-2021-0003).

\subsection*{Conflict of interest}

The author declares that he has no conflicts of interest.

\subsection*{Open access}

This article is licensed under a Creative Commons Attribution 
4.0 International License, which permits use, sharing,
adaptation, distribution and reproduction in any medium or
format, as long as you give appropriate credit to the original
author(s) and the source, provide a link to the Creative Commons 
license, and indicate if changes were made. The images
or other third party material in this article are included in the
article’s Creative Commons license, unless indicated otherwise
in a credit line to the material. If material is not included
in the article’s Creative Commons license and your intended
use is not permitted by statutory regulation or exceeds the
permitted use, you will need to obtain permission directly
from the copyright holder. To view a copy of this license, visit\\
\url{http://creativecommons.org/licenses/by/4.0/} .


\begin{thebibliography}{99}

\bibitem{opt1} A. L. Berkhoer and V. E. Zakharov, 
Sov. Phys. JETP {\bf 31}, 486 (1970).

\bibitem{opt2}  Y. Kivshar and G. P. Agrawal,
{\it Optical Solitons: From Fibers to Photonic Crystals}, 
(Academic, San Diego, CA, 2003).

\bibitem{opt3} V. E. Zakharov and S. Wabnitz, 
{\it Optical Solitons: Theoretical Challenges and Industrial Perspectives}
(Springer, Berlin, 1999).

\bibitem{opt4} B. A. Malomed, {\it Multidimensional Solitons} 
(AIP, Melville, N. Y., 2022)\\ 
\url{https://doi.org/10.1063/9780735425118}

\bibitem{mix1} Tin-Lun Ho and V. B. Shenoy, Phys. Rev. Lett. {\bf 77}, 3276 (1996).

\bibitem{mix2} H. Pu and N. P. Bigelow, Phys. Rev. Lett. {\bf 80}, 1130 (1998).

\bibitem{mix3} B. P. Anderson, P. C. Haljan, C. E. Wieman, and E. A. Cornell,
Phys. Rev. Lett. {\bf 85}, 2857 (2000).

\bibitem{mix4} S. Coen and M. Haelterman, Phys. Rev. Lett. {\bf 87}, 140401 (2001).

\bibitem{mix5} G. Modugno, M. Modugno, F. Riboli, G. Roati, and M. Inguscio, 
Phys. Rev. Lett. {\bf 89}, 190404 (2002).

\bibitem{sep} E. Timmermans, Phys. Rev. Lett. {\bf 81}, 5718 (1998).

\bibitem{AC1998} P. Ao and S. T. Chui, Phys. Rev. A {\bf 58}, 4836 (1998).

\bibitem{tension} B. Van Schaeybroeck, Phys. Rev. A {\bf 78}, 023624 (2008).

\bibitem{bubbles} K. Sasaki, N. Suzuki, and H. Saito,
Phys. Rev. A {\bf 83}, 033602 (2011).

\bibitem{KHI1} H. Takeuchi, N. Suzuki, K. Kasamatsu, H. Saito, and M. Tsubota,
Phys. Rev. B {\bf 81}, 094517 (2010).

\bibitem{KHI2} N. Suzuki, H. Takeuchi, K. Kasamatsu, M. Tsubota, and H. Saito,
Phys. Rev. A {\bf 82}, 063604 (2010).

\bibitem{KHI3} H. Kokubo, K. Kasamatsu, and H. Takeuchi,
Phys. Rev. A {\bf 104}, 023312 (2021). 
 
\bibitem{RTI1}  K. Sasaki, N. Suzuki, D. Akamatsu, and H. Saito,
Phys. Rev. A {\bf 80}, 063611 (2009).

\bibitem{RTI2} S. Gautam and D. Angom,  Phys. Rev. A {\bf 81}, 053616 (2010).

\bibitem{RTI3} T. Kadokura, T. Aioi, K. Sasaki, T. Kishimoto, and H. Saito,
Phys. Rev. A {\bf 85}, 013602 (2012).

\bibitem {capillary} K. Sasaki, N. Suzuki, and H. Saito,
Phys. Rev. A {\bf 83}, 053606 (2011).

\bibitem {param_inst-1} D. Kobyakov, V. Bychkov, E. Lundh, A. Bezett, 
and M. Marklund, Phys. Rev. A {\bf 86}, 023614 (2012).

\bibitem {param_inst-2}  D. K. Maity, K. Mukherjee, S. I. Mistakidis, S. Das, 
P. G. Kevrekidis, S. Majumder, and P. Schmelcher, 
Phys. Rev. A {\bf 102}, 033320, (2020).

\bibitem{mix-sheet-1} K. Kasamatsu, M. Tsubota, and M. Ueda,
Phys. Rev. Lett. {\bf 91}, 150406 (2003).

\bibitem{mix-sheet-2} K. Kasamatsu and M. Tsubota,
Phys. Rev. A {\bf 79}, 023606 (2009).

\bibitem{topo_defects} P. Mason and A. Aftalion, 
Phys. Rev. A {\bf 84}, 033611 (2011).

\bibitem{vortex-mol} K. Kasamatsu, M. Tsubota, and M. Ueda, 
Phys. Rev. Lett. {\bf 93}, 250406 (2004).

\bibitem{wall-annih-1} H. Takeuchi, K. Kasamatsu, M. Tsubota, and M. Nitta,
Phys. Rev. Lett. {\bf 109}, 245301 (2012).

\bibitem{wall-annih-2} M. Nitta, K. Kasamatsu, M. Tsubota, and H. Takeuchi,
Phys. Rev. A {\bf 85}, 053639 (2012).

\bibitem{vortex-wall} K. Kasamatsu, H. Takeuchi, M. Tsubota, and M. Nitta,
Phys. Rev. A {\bf 88}, 013620 (2013).

\bibitem{R2021-2} V. P. Ruban, JETP Lett. {\bf 113}, 814 (2021).

\bibitem{VB1} K. J. H. Law, P. G. Kevrekidis, and L. S. Tuckerman,
Phys. Rev. Lett. {\bf 105}, 160405 (2010); 
{\it Erratum}, Phys. Rev. Lett. {\bf 106}, 199903 (2011).

\bibitem{VB2} M. Pola, J. Stockhofe, P. Schmelcher, and P. G. Kevrekidis,
Phys. Rev. A {\bf 86}, 053601 (2012).

\bibitem{VB3} S. Hayashi, M. Tsubota, and H. Takeuchi,
Phys. Rev. A {\bf 87}, 063628 (2013).

\bibitem{m-vort-2D-1} A. Richaud, V. Penna, R. Mayol, and M. Guilleumas,
Phys. Rev. A {\bf 101}, 013630 (2020).

\bibitem{m-vort-2D-2} A. Richaud, V. Penna, and A. L. Fetter,
Phys. Rev. A {\bf 103}, 023311 (2021).

\bibitem{R2021} V. P. Ruban, JETP Lett. {\bf 113}, 532 (2021).

\bibitem{RWTK2022} V. P. Ruban, W. Wang, C. Ticknor, and P. G. Kevrekidis,
Phys. Rev. A {\bf 105}, 013319 (2022).

\bibitem{R2022} V. P. Ruban, JETP Lett. {\bf 115}, 415 (2022).

\bibitem{PDWopp1} V. E. Zakharov and A. V. Mikhailov, 
JETP Lett. {\bf 45}, 349 (1987).

\bibitem{PDWopp2} S. Pitois, G. Millot, and S. Wabnitz,
Phys. Rev. Lett. {\bf 81}, 1409 (1998). 

\bibitem{PDW1} M. Haelterman and A. P. Sheppard,
Phys. Rev. E {\bf 49}, 3389 (1994).

\bibitem{PDW2} M. Haelterman and A. P. Sheppard,
Phys. Rev. E {\bf 49}, 4512 (1994).

\bibitem{PDW3} A. P. Sheppard and M. Haelterman,
Opt. Lett. {\bf 19}, 859 (1994).

\bibitem{PDW4} N. Dror, B. A. Malomed, and J. Zeng,
Phys. Rev. E {\bf 84}, 046602 (2011).

\bibitem{PDW5} Yu. S. Kivhsar and B. Luther-Davies,
Phys. Rep. {\bf 298}, 81 (1998).

\bibitem{OVS1} A. H. Carlsson, J. N. Malmberg, D. Anderson, M. Lisak, 
E. A. Ostrovskaya, T. J. Alexander, and Yu. S. Kivshar,
Opt. Lett. {\bf 25}, 660 (2000).

\bibitem{OVS2} A. S. Desyatnikov, L. Torner, and Yu. S. Kivshar,
Progress in Optics {\bf 47}, 291 (2005).

\bibitem{RA2000}
S. Raghavan and G. P. Agrawal, Opt. Communications {\bf 180}, 377 (2000).

\bibitem{GRIN1} S. Longhi, Opt. Lett. {\bf 28}, 2363 (2003). 

\bibitem{GRIN2} A. Mafi, 
J. Lightwave Technology {\bf 30}, 2803 (2012).

\bibitem{GRIN3} C. M. Arabi, A. Kudlinski, A. Mussot, and M. Conforti,
Phys. Rev. A {\bf 97}, 023803 (2018).

\bibitem{GRIN4} T. Mayteevarunyoo, B. A. Malomed, and D. V. Skryabin,
J. Opt. {\bf 23}, 015501 (2020).

\bibitem{GRIN5} L. G. Wright, F. O. Wu, D. N. Christodoulides, 
and F. W. Wise, Nat. Phys. {\bf 18}, 1018 (2022).
 
\bibitem{video1} \url{http://home.itp.ac.ru/~ruban/27DEC2022/w1.avi}

\bibitem{video1a} \url{http://home.itp.ac.ru/~ruban/27DEC2022/w1a.avi}

\bibitem{video2} \url{http://home.itp.ac.ru/~ruban/27DEC2022/w2.avi}

\bibitem{video3} \url{http://home.itp.ac.ru/~ruban/27DEC2022/w3.avi}

\bibitem{Necklace-Ring-Vector-Solitons}
A. S. Desyatnikov and Yu. S. Kivshar,
Phys. Rev. Lett. {\bf 87}, 033901 (2001).

\bibitem{Polarization_Shaping} F. Bouchard, H. Larocque, A. M. Yao, C. Travis,
I. De Leon, A. Rubano, E. Karimi, G.-L. Oppo, and R. W. Boyd,
Phys. Rev. Lett. {\bf 117}, 233903 (2016).

\end{thebibliography}
\end{document}